\documentclass[aps,prd,twocolumn,showpacs,groupedaddress,10pt,nofootinbib]{revtex4}
\usepackage{graphicx}
\usepackage{amssymb}
\usepackage{amsmath}
\usepackage{amsfonts}
\begin{document}

\title{Quintom phase-space: beyond the exponential potential}

\author{Genly Leon}
\email{genly.leon@ucv.cl}
\affiliation{Instituto de F\'{\i}sica, Pontificia Universidad  Cat\'olica de Valpara\'{\i}so, Casilla 4950, Valpara\'{\i}so, Chile}
\author{Yoelsy Leyva}
\email{yoelsy.leyva@fisica.ugto.mx} \affiliation{Departamento de
F\'{\i}sica, DCI, Universidad de Guanajuato-Campus Le\'on,
 C.P. 37150, Le\'on, Guanajuato, M\'exico}

\author{J. Socorro}
\email{socorro@fisica.ugto.mx}
 \affiliation{Departamento de  F\'{\i}sica, DCI, Universidad de Guanajuato-Campus Le\'on,
 C.P. 37150, Le\'on, Guanajuato, M\'exico}

\begin{abstract}
We investigate the phase-space structure of the quintom dark energy paradigm in the framework of spatially flat and
 homogeneous universe. Considering arbitrary decoupled potentials, we find certain general conditions under which the phantom dominated solution is late time attractor, generalizing previous results found for the case of exponential potential.  Center Manifold Theory is employed to obtain sufficient conditions for the instability of de Sitter solution either with phantom or quintessence potential dominance. 
\end{abstract}

\maketitle

\section{Introduction}
Recent cosmological observations point to an strong evidence for an spatially flat and accelerated expanding universe
\cite{Riess2011, Komatsu2011, Reid2010}. Despite the great agreement of observations with the concordance model
\cite{Li2011} \footnote{The Cosmological Constant Model.}, it is a fact that quintom model, whose Equation of State
(EoS) can cross the cosmological constant barrier $w=-1$, is not exclude by observations \cite{Alam2004a,Feng2005,Huterer2005,Nesseris2007,Jassal2010,Novosyadlyj2012,Novosyadlyj2012a,Hinshaw2012}. A popular way
to realize a viable quintom model and, at the same time, avoid the restrictions imposed by the \textit{No-Go Theorem} \cite{Vikman2005,Hu2005, Caldwell2005, Xia2008, Cai2010}
is the introduction of extra degrees of freedom\footnote{The only way to realize the crossing
without any ghosts and gradient instabilities in standard gravity and with one single scalar degree of freedom was obtained in \cite{Deffayet2010}. 
}. Following
this recipe, the simple quintom paradigm requires a canonical quintessence scalar
field $\sigma$ and simultaneously a phantom scalar field $\phi$ where the effective potential can be of arbitrary form,
while the two components can be either coupled \cite{Zhang2006} or decoupled \cite{Feng2005, Guo2005}. 

The properties of the quintom models have been studied from different points of view. Among them, the phase space studies,
using the dynamical systems tools, are very useful in order to analyze the asymptotic behavior of the model. In quintom models
 this program have been carried out in \cite{Guo2005, Zhang2006, Lazkoz2006, Lazkoz2007, Setare2009, Setare2008a, Setare2009c, Cai2010}.
 In \cite{Guo2005} the decoupled case between the canonical and phantom field with an exponential potential is studied shown that
 the phantom-dominated scaling solution is the unique late-time attractor. In \cite{Zhang2006} the potential considers the interaction
  between the fields and shows that in the absence of interactions, the solution dominated by the phantom field should be the attractor
  of the system  and the interaction does not affect its attractor behavior. This result is correct only in the case  in which the
  existence of the phantom phase excludes the existence scaling attractors \cite{Lazkoz2006}. Some of these
results were extended in \cite{Lazkoz2007} for arbitrary potentials. In \cite{Setare2009c} the authors showed that all quintom models
 with nearly flat potentials converge to a single expression for EoS of dark energy, in addition, the necessary conditions for the
  determination of the direction of the $w=-1$ crossing was found.

The aim of this paper is to extend the study of Refs. \cite{Guo2005,
Lazkoz2006, Lazkoz2007, Setare2009, Cai2010} -investigation of the
dynamics of quintom cosmology- to include a wide variety of
potential beyond the exponential potential without interaction
between the fields, all of them can be constructed using the Bohm
formalism \cite{Guzman2007, Socorro2010, Bohm1952} of the quantum
mechanics under the integral systems premise, which is known as
quantum potential approach.  This approach makes it possible to
identify trajectories associated with the wave function of the
universe \cite{Guzman2007} when we choose the superpotential
function as the momenta associated to the coordinate field $q$. This
investigation was undertaken within the framework of the
minisuperspace approximation to quantum theory when we investigate
the dynamics of only a finite number of models. Here we make use of
the dynamical systems tools to obtain useful information about the
asymptotic properties of the model. In order to be able to analyze
self-interaction potentials beyond the exponential one, we rely on
the method introduced in Ref. \cite{Fang2009} in the context of
quintessence models and that have been generalized to several cosmological contexts
like: Randall-Sundrum II and DGP branes \cite{Leyva2009, Escobar2012a,
Escobar2012e}, Scalar Field Dark Matter models \cite{Matos2009},
tachyon and phantom fields \cite{Quiros2010, Fang2010a,
Farajollahi2011} and loop quantum gravity \cite{Xiao2011j}.

The plan of the paper is as follow: in section \ref{ss1} we introduce the quintom model for arbitrary potentials and in section \ref{ss2}
 we build the corresponding autonomous system. The results of the study of the corresponding critical points,
 their stability properties and the physical discussion are shown in section \ref{s2}. The section \ref{ss4} is devoted to conclusions. Finally,  we include in the two appendices \ref{apen1} and \ref{apen2} the center manifold calculation of the solutions dominated by either the phantom or quintessence potential. 

\section{The model}\label{ss1}
The starting action of our model, containing the canonical field $\sigma$ and the phantom field $\phi$, is \cite{Feng2005, Guo2005, Zhang2006}:

\begin{eqnarray}\label{action}
S=\int d^{4}x\sqrt{-g}\left(\frac{1}{2}R-\frac{1}{2}g^{\mu\nu}\partial_{\mu}\sigma\partial_{\nu}\sigma+V_{\sigma}(\sigma) + \right. \nonumber\\
    \left. + \frac{1}{2}g^{\mu\nu}\partial_{\mu}\phi\partial_{\nu}\phi+V_{\phi}(\phi)\right),
\end{eqnarray}

where we used natural units ($8 \pi G=1$) and $V_{\sigma}(\sigma)$ and $V_{\phi}(\phi)$ are respectively the self interactions
potential of the quintessence and phantom fields. 

From this action the Friedmann equations for a flat geometry reads \cite{Guo2005, Zhang2006}:
\begin{equation}\label{F1}
    H^{2}=\frac{1}{3}\left( \frac{\dot{\sigma}^{2}}{2}+V_{\sigma}(\sigma)-\frac{\dot{\phi}^{2}}{2}+V_{\phi}(\phi)\right)
\end{equation}
\begin{equation}\label{F2}
    \dot{H}=-\frac{1}{2}\left(\dot{\sigma}^{2}-\dot{\phi}^{2} \right)
\end{equation} where 
where $H=\frac{\dot{a}}{a}$ is the Hubble parameter and the dot denotes derivative with respect the time. 

The evolution of the quintessence and phantom field are:
\begin{equation}\label{KG1}
    \ddot{\sigma}+3H\dot{\sigma}+V_{\sigma}'(\sigma)=0
\end{equation}
\begin{equation}\label{KG2}
    \ddot{\phi}+3H\dot{\phi}- V_{\phi}'(\phi)=0,
\end{equation} where the coma denotes the derivative of a function with respect to their argument. 

Additionally we can introduce the total energy density and pressure as:
\begin{equation}
    \rho_{DE}=\rho_{\sigma}+\rho_{\phi},\;\;\;p_{DE}=p_{\sigma}+p_{\phi}
\end{equation}
where
\begin{equation}
    \rho_{\sigma}=\frac{\dot{\sigma}^2}{2}+V_{\sigma}(\sigma),\;\;\;\rho_{\phi}=-\frac{\dot{\phi}^2}{2}+V_{\phi}(\phi)
\end{equation}
\begin{equation}
    p_{\sigma}=\frac{\dot{\sigma}^2}{2}-V_{\sigma}(\sigma),\;\;\;p_{\phi}=-\frac{\dot{\phi}^2}{2}-V_{\phi}(\phi)
\end{equation}
and its equation of state parameter is given by
\begin{equation}\label{weff}
    w_{eff}=\frac{p_{\sigma}+p{\phi}}{\rho_{\sigma}+\rho_{\phi}}=\frac{\dot{\sigma}^{2}-\dot{\phi}^{2}-2V_{\sigma}(\sigma)-2V_{\phi}(\phi)}{\dot{\sigma}^{2}-\dot{\phi}^{2}+2V_{\sigma}(\sigma)+2V_{\phi}(\phi)}
\end{equation}
and 
\begin{equation}\label{omegaaa}
    \Omega_{\sigma}=\frac{\rho_{\sigma}}{\rho_{DE}},\;\;\Omega_{\phi}=\frac{\rho_{\phi}}{\rho_{DE}}
\end{equation}
\begin{equation}
    \Omega_{\sigma}+ \Omega_{\phi}=1
\end{equation}
are the the individual and total dimensionless densities parameters. 

\section{The autonomous system}\label{ss2}
In order to study the dynamical properties of the system
(\ref{F1}-\ref{KG2}) we introduce the following dimensionless phase
space variables to build an autonomous system \cite{Copeland1998, Chen2009}:
\begin{equation}\label{nv1}
x_{\sigma}=\frac{\dot{\sigma}}{\sqrt{6}H}, \\
x_{\phi}=\frac{\dot{\phi}}{\sqrt{6}H}, \\
y_{\sigma}=\frac{\sqrt{V_{\sigma}(\sigma)}}{\sqrt{3}H}, \\
\end{equation}
\begin{equation}
\lambda_{\sigma}=-\frac{V_{\sigma}'(\sigma)}{V_{\sigma}(\sigma)}, \\
\lambda_{\phi}=-\frac{V_{\phi}'(\phi)}{V_{\phi}(\phi)}, \\
\end{equation}

Notice that the phase space variables $\lambda_{\sigma}$ and
$\lambda_{\phi}$ are sensitive of the kind of self interactions
potential chosen for quintessence and phantom component,
respectively and are introduced in order to be able to study
arbitrary potentials. Applying the above dimensionless variables to
the system (\ref{F1}-\ref{KG2}) we obtain the following autonomous
system:

\begin{eqnarray}
 \frac{d x_{\sigma}}{dN}&=& -3 x_{\sigma} \left(1+x_\phi^2-x_\sigma^2 \right)+\sqrt{\frac{3}{2}} y_{\sigma}^2 \lambda_\sigma \label{auto1}\\
 \frac{d x_{\phi}}{dN}&=& -3 x_{\phi} \left(1+x_\phi^2-x_\sigma^2 \right)+\nonumber\\&&-\sqrt{\frac{3}{2}}\left(1+x_\phi^2-x_\sigma^2-y_\sigma^2\right) \lambda_\phi \label{auto2}\\
\frac{d y_{\sigma}}{dN} &=&\frac{1}{2}y_\sigma \left(6x_\sigma^2-\sqrt{6} x_\sigma
   \lambda_\sigma -6 x_\phi^2\right) \label{auto3}\\
 \frac{d \lambda_{\sigma}}{dN}&=& -\sqrt{6} x_{\sigma} f(\lambda_\sigma) \label{auto4}\\
\frac{d \lambda_{\phi}}{dN} &=& -\sqrt{6}x_\phi g(\lambda_\phi)\label{auto5}
\end{eqnarray}
where $N=\ln a$ is the number of e-foldings and  $f(\lambda_\sigma)= \lambda_\sigma
^2(\Gamma_{\sigma}-1)$  and $ g(\lambda_\phi)= \lambda_\phi
^2(\Gamma_{\phi}-1) $ where:
\begin{equation}
\Gamma_{\sigma}=\frac{V_{\sigma}(\sigma)V_{\sigma}''(\sigma)}{(V_{\sigma}'(\sigma))^{2}},
\;\;\; \Gamma_{\phi}=\frac{V_{\phi}(\phi)V_{\phi}''(\phi)}{(V_{\phi}'(\phi))^{2}}
\end{equation}
In order to get from the autonomous equation (\ref{auto1}-\ref{auto5})  a closed system of ordinary differential equation
we have assumed that the funtions $\Gamma_{\sigma}$ and $\Gamma_{\phi}$ can be written as a function of the
variables $\lambda_{\sigma}\in\mathbb{R}$ and $\lambda_{\phi}\in\mathbb{R}$ respectively \cite{Fang2009}.

The phase space for the autonomous dynamical system driven by de evolutions of Eqs. (\ref{auto1}-\ref{auto5}) can be defined as follows:
\begin{eqnarray}
    \Psi=\{(x_{\sigma},x_{\phi},y_{\sigma}):y_{\sigma}\geq 0, x_\sigma^2-x_\phi^2+y_\sigma^2\leq 1 \}\times\nonumber\\
                \times\{(\lambda_{\sigma},\lambda_{\phi})\in\mathbb{R}^{2} \}
\end{eqnarray}

With the aim of explain the physical significance of the critical points of the autonomous system (\ref{auto1}-\ref{auto5}) we need to obtain the relevant cosmological parameters in terms of the dimensionless phase space variables (\ref{nv1}). Following this, the cosmological parameter (\ref{weff}) and (\ref{omegaaa}) can be expressed as
\begin{equation}
 w_{eff}=-1+2x_{\sigma}^2-2x_{\phi}^2
\end{equation}
\begin{equation}
 \Omega_{\sigma}=x_{\sigma}^2+y_{\sigma}^2,\;\;\ \Omega_{\phi}=1-x_{\sigma}^2-y_{\sigma}^2,
\end{equation}
while the deceleration parameter becomes
\begin{equation}
 q=-\left[1+\frac{\dot{H}}{H^2}\right]=-1+3x_{\sigma}^2-3x_{\phi}^2.
\end{equation}

\section{Critical points and stability}\label{s2}
The critical points of the system (\ref{auto1}-\ref{auto2}) are
summarized in Table \ref{tab1}. The eigenvalues of the corresponding
Jacobian matrices are show in Table \ref{tab2}. In both cases
$\lambda_{\sigma}^{\ast}$ and $\lambda_{\phi}^{\ast}$ are the values
which makes the functions
$f(\lambda_{\sigma})=\lambda_\sigma^2\left(\Gamma_{\sigma}-1\right)$
and $g(\lambda_{\phi})=\lambda_\phi^2\left(\Gamma_{\phi}-1\right)$
vanish respectively. 

\begin{table*}\caption[crit]{Properties of the critical points for the
autonomous system (\ref{auto1}-\ref{auto5})}
\begin{tabular}
{l c c c c c c c c c c}
\hline\hline\\[-0.3cm]
$Label$&$x_{\sigma}$&$y_{\sigma}$&$x_{\phi}$&$\lambda_{\sigma}
$&$\lambda_{\phi}$&Existence&$\Omega_{\sigma}$&$\Omega_{\phi}
$&$q$&$w_{eff}$\\
\hline\\[-0.2cm]

$P_1^\pm$ &$0$  &$0$  &$\pm i$  &$\lambda_{\sigma}$  &
$\lambda_{\phi}^{\ast}$ &  Non real   &$0$  &$1$& $2$ & $1$\\[0.2cm]

$P_2^\pm$ &$\pm1$  & $0$  &$0$  &$\lambda_{\sigma}^{\ast}$
&$\lambda_{\phi}$&   Always      &$1$ &$0$ &$2$ &$1$ \\[0.2cm]

$P_3^\pm$ &$\pm\sqrt{1+x_\phi^2}$  &$0$  & $x_\phi$ &$\lambda_{\sigma}^{\ast}$ &
$\lambda_{\phi}^{\ast}$&   \textquotedblright      &$1+x_{\phi}^{2}$ &$-x_{\phi}^{2}$
&$2$ &$1$ \\[0.2cm]

$P_{4}$ &$\frac{\lambda_{\sigma}^{\ast}}{\sqrt{6}}$
&$\sqrt{1-\frac{(\lambda_{\sigma}^{\ast})^{2}}{6}}$&$0$
&$\lambda_{\sigma}^{\ast}$ &$\lambda_{\phi}$&  $-\sqrt{6}\leq \lambda_{\sigma}^{\ast} \leq \sqrt{6}$       &$1$ &$0$
&$-1+\frac{(\lambda_{\sigma}^{\ast})^{2}}{2}$ &
$-1+\frac{(\lambda_{\sigma}^{\ast})^{2}}{3}$  \\[0.2cm]

$P_5$ & $0$ & $0$ & $0$ & $\lambda_{\sigma}$ &
$0$& Always & $0$  & $1$ &$-1$ & $-1$ \\[0.2cm]

$P_6$ &$0$  &$1$  &$0$  &$0$  &$\lambda_{\phi}$  & \textquotedblright & $1$ &
$0$ & $-1$ &$-1$ \\[0.2cm]

$P_{7}$ &$0$  &$y_{\sigma}$  &$0$  & $0$
&$0$ &  $0< y_{\sigma}< 1$     &$y_{\sigma}^{2}$  &$1-y_{\sigma}^{2}$ &
$-1$ &$-1$\\ [0.2cm]

$P_{8}$&$0$  &$0$ &$-\frac{\lambda_{\phi}^{\ast}}{\sqrt{6}}$
&$\lambda_{\sigma}$ & $\lambda_{\phi}^{\ast}$ &   $\lambda_{\sigma}
\in \mathbb{R}$ &$0$ &$1$
&$-1-\frac{(\lambda_{\phi}^{\ast})^{2}}{2}$
&$-1-\frac{(\lambda_{\phi}^{\ast})^{2}}{3}$ \\ [0.2cm] \hline \hline
\end{tabular}\label{tab1}
\\ [0.2cm]
\end{table*}

\begin{table*}\caption[crit2]{Eigenvalues of the linear perturbation matrix associated to each of the critical points displayed in Table \ref{tab1}}
\begin{tabular}
{l c c c c c c c c c c}
\hline\hline\\[-0.3cm]
$Label$&$m_1$&$m_2$&$m_3$&$m_4$&$m_5$\\
\hline\\[-0.2cm]

$P_1^\pm$ & $3$ & $0$ & $0$ & $\mp {i}\sqrt{6}g'(\lambda_{\phi}^{\ast})$  & $6\mp{ i}\sqrt{6}\lambda_{\phi}^{\ast}$\\[0.2cm]

$P_2^\pm$ & $6$ & $0$ & $0$ & $\mp\sqrt{6}f'(\lambda_{\sigma}^{\ast})$ & $3\mp\sqrt{\frac{3}{2}}\lambda_\sigma^{\ast}$\\[0.2cm]

$P_3^\pm$ & $0$ & $-\sqrt{6} g'(\lambda_{\phi}^{\ast}) {x_\phi}$&$\mp\sqrt{6} {f'(\lambda_{\sigma}^{\ast})}
   \sqrt{x_\phi^2+1}$&$3\mp\sqrt{\frac{3}{2}} \sqrt{x_\phi^2+1}
   \lambda_\sigma ^{\ast}$& $6-\sqrt{6} x_\phi \lambda_\phi ^{\ast}$ \\[0.2cm]

$P_4$ & $0$ &$-{f'(\lambda_{\sigma}^{\ast})} \lambda \sigma ^{\ast}$ & $\left(\lambda \sigma
   ^{\ast}\right)^2$ &$\frac{1}{2} \left(\left(\lambda \sigma
   ^{\ast}\right)^2-6\right)$ &$\frac{1}{2} \left(\left(\lambda \sigma
   ^{\ast}\right)^2-6\right)$\\ [0.2cm]

$P_5$ & $-3$ & $0$ & $0$ & $-\frac{3}{2}\left(1+\sqrt{1+\frac{4}{3}g(0)}\right)$ & $-\frac{3}{2}\left(1-\sqrt{1+\frac{4}{3}g(0)}\right)$\\[0.2cm]

$P_6$ & $-3$ & $0$ & $0$ & $-\frac{3}{2}\left(1+\sqrt{1-\frac{4}{3}f(0)}\right)$ & $-\frac{3}{2}\left(1-\sqrt{1-\frac{4}{3}f(0)}\right)$\\[0.2cm]

$P_7$ & $0$ & $\frac{1}{2} \left(-\sqrt{9-12 f(0) y_\sigma^2}-3\right)$ & $\frac{1}{2} \left(\sqrt{9-12 f(0) y_\sigma^2}-3\right)$ & $\frac{1}{2} \left(-\sqrt{9-12 g(0) \left(y_\sigma^2-1\right)}-3\right)$ & $\frac{1}{2} \left(\sqrt{9-12 g(0)
   \left(y_\sigma^2-1\right)}-3\right)$\\[0.2cm]

$P_8$ & $0$ &${g'(\lambda_{\phi}^{\ast})} \lambda_\phi ^{\ast}$& $-\frac{1}{2} \left(\lambda_\phi
   ^{\ast}\right)^2$&$-\frac{1}{2} \left(\left(\lambda_\phi
   ^{\ast}\right)^2+6\right)$&$-\frac{1}{2} \left(\left(\lambda_\phi
   ^{\ast}\right)^2+6\right)$\\ [0.2cm]


\hline \hline\\[-0.3cm]
\end{tabular}\label{tab2}
\end{table*}
As we see from Table \ref{tab1}, the points $P_1^\pm$ do not exist in the strict sense ($x_\phi$ is purely imaginary at the fixed points).
Point $P_5$ is associated with a combination of a phantom potential whose first $\phi$-derivative vanishes at some/several point/points, i.e.,
$\lambda_\phi=0$ (this case include the exponential potential whose $\phi$-derivative at any order vanish everywhere) and an arbitrary self
interaction potential for the quintessence component (arbitrary value of $\lambda_{\sigma}$). Point $P_6$ is associated with a combination
 of a quintessence potential whose first $\sigma$-derivative vanishes at some/several point/points, i.e., $\lambda_\sigma=0$
 (this case include the exponential potential whose $\sigma$-derivative at any order vanish everywhere) and an arbitrary self
 interaction potential for the phantom component (arbitrary value of $\lambda_{\phi}$). Point $P_7$  is associated with a
 combination of a phantom potential whose first $\phi$-derivative vanishes at some/several point/points, i.e., $\lambda_\phi=0$
 (this case include the exponential potential whose $\phi$-derivative at any order vanish everywhere) and a self interaction
 potential for the quintessence component whose first $\sigma$-derivative vanishes also at some/several point/points, i.e., $\lambda_\phi=0.$ 
It is worth noticing that the existence of points  $P_2^\pm$, $P_3^\pm$, $P_4$,  $P_8$ and $P_9$  depends of the concrete form
of the potential. From the table of the eigenvalues, notice, besides, that all the points belongs to nonhyperbolic sets of
critical point with a least one null  eigenvalue.

\subsection{Stability of the critical points}

Although  all these critical points are shown in the Tables \ref{tab1} here we have summarized their basic properties:
\begin{itemize}
    \item $P_{1}^\pm$, $P_{2}^\pm$ and $P_{3}^\pm$ correspond to a solution dominated by the kinetic energy of the scalar fields
    (stiff fluid solution: $q=2$ and $\omega=1$). The exact dynamical behavior differs for each points.  $P_1^\pm$ corresponds to
    a phantom kinetic energy dominated  ($\Omega_{\sigma}=0$ and $\Omega_{\phi}=1$). However, these points have a purely imaginary
    value of $x_\phi,$ thus, they do not exists in the strict sense. They have a three$-$dimensional center subspace and a two-dimensional
    unstable manifold ($m_1=3>0,\; \Re(m_5)=6>0$). Thus they cannot be late-time attractors. $P_3^\pm$ represents an scaling regimen
    between the kinetics energies of the quintessence and phantom fields ($\Omega_{\sigma}=1+x_{\phi}^{2}$ and $\Omega_{\phi}=-x_{\phi}^{2}$).
    These points depend of the form of the potentials and under certain conditions they have a four dimensional unstable subspace which could
    correspond to the past attractor. However, this point is unphysical since $\Omega_{\phi}<0$ . $P_{2}^\pm$ is dominated by the
    quintessence kinetic term ($\Omega_{\sigma}=1$ and $\Omega_{\phi}=0$). Since they are non-hyperbolic  due to the existence of
    two null eigenvalues,  we are not able to extract information about their stability by  using the standard tools of the linear
    dynamical analysis. However, since these points seems to be  particular cases of $P_3^\pm,$ they should share the same dynamical
    behavior. Because all of these points are nonhyperbolic, as we notice before, we cannot rely on the standard  linear
    dynamical systems analysis for deducing their stability. Thus, we need to rely our analysis on numerical inspection of
     the phase portrait for specific potentials or use more sophisticated techniques like Center Manifold theory.
    \item $P_4$ is an scaling solution between the kinetic and the potential energy of the quintessence component of dark energy. This solution in sensitive to the explicit form of the potential. This is always a saddle equilibrium point in the phase space since $m_2=(\lambda_{\sigma}^{\ast})^{2}$ and $m_4=\frac{1}{2}((\lambda_{\sigma}^{\ast})^{2}-6)$ are of opposite sign in the existence region of  this point. It represents an accelerated solution for a potential $V_\sigma(\sigma)$ whose function $f(\lambda_{\sigma})$ vanish for $\lambda_{\sigma}=\lambda_{\sigma}^{\ast}$ in the interval $-\sqrt{2}<\lambda_{\sigma}^{\ast}<\sqrt{2}$,  leading to a $-1\leq w_{eff}<-1/3$. When $\lambda_{\sigma}^{\ast}=0$ the critical point $P_4$ becomes in $P_6$. In the regions $-\sqrt{6}\leq\lambda_{\sigma}^{\ast}\leq-\sqrt{2}$ or $\sqrt{2}\leq\lambda_{\sigma}^{\ast}\leq\sqrt{6}$, the critical point $P_4$ represents a non-accelerated phase. A very interesting issue of this critical point appears when, for an specific form of
the quintessence potential, $\lambda_{\sigma}^{\ast}=\pm\sqrt{3}$, driving to $w_{eff}=0$. This means that the quintessence field is able to mimic the dark matter behavior.
    \item $P_5$, $P_6$ and $P_7$ represents solutions dominated by the potential energies of the potentials (all of them represent de Sitter solutions: $q=-1$ and $w_{eff}=-1$). Once again the exact dynamical nature differs from one point to the other: $P_5$ is dominated by the potential energy of the phantom component ($\Omega_{\sigma}=0$ and $\Omega_{\phi}=1$). Because of the existence of two null eigenvalues  is not possible to conclude about its dynamics. However it has a three-dimensional stable manifold for $g(0)<0$ (in the interval $g(0)<-\frac{3}{4}$ it has to  complex conjugated eigenvalues with negative real parts). In this cases it is worthy to analyze its stability using the center manifold theory. $P_6$ is a critical point dominated by the quintessence potential energy term ($\Omega_{\sigma}=1$ and $\Omega_{\phi}=0$), despite its nonhyperbolicity, it has  three-dimensional stable manifold for $f(0)>0$ (in the case $f(0)>\frac{3}{4}$ it has to  complex conjugated eigenvalues with
negative real parts), thus, it is worthy to analyze its stability using the center manifold theory.  $P_7$ denotes a segment (curve) of non-isolated fixed points, representing a scaling regimen between the quintessence and phantom potential ($\Omega_{\sigma}=y_{\sigma}^{2}$ and $\Omega_{\phi}=1- y_{\sigma}^{2}$).  The existence of one non-zero eigenvalue is due to the fact that it is a curve of fixed points. As an invariant set of non-isolated singular points it is normally-hyperbolic, since the eigenvector associated to the zero eigenvalue, $(0,0,1,0,0)^T,$ is tangent to the curve. Thus its stability is determined by the sign of the remaining non-null eigenvalues. Hence, it is stable for $0<y_\sigma<1,\;  f(0)>0, \; g(0)<0$ or a saddle otherwise.
\item $P_8$ is a line of fixed points parameterized by $\lambda_\sigma\in\mathbb{R}$. The existence of one non-zero eigenvalue is due to the fact that it is a curve of fixed points. As an invariant set of non-isolated singular points it is normally-hyperbolic, since the eigenvector associated to the zero eigenvalue, $(0,0,0,1,0)^T,$ is tangent to the curve. Thus its stability is determined by the sign of the remaining non-null eigenvalues.  From table \ref{tab2} follows that $P_8$ admits a four dimensional stable subspace provided $g'(\lambda_{\phi}^{\ast}) \lambda_{\phi}^{\ast}<0$, thus, the invariant curve is stable. It represents  accelerated solutions dominated by the phantom potential  providing a crossing through the phantom divide ($\Omega_{\sigma}=0$ and $\Omega_{\phi}=1$). For every value of $\lambda_{\phi}^{\ast}$ this point provide the typical superaccelerated expansion of quintom paradigm ($w=-1-\frac{(\lambda_{\phi}^{\ast})^{2}}{3}$) the only exception occurs when $\lambda_{\phi}^{\ast}=0$
recovering the behavior of the de Sitter solution  $P_5$  ($\omega=-1$). This line of critical point corresponds to the stable point $P$ in \cite{Guo2005} and B in \cite{Cai2010} (phantom dominated solution). Summarizing, the line $P_8$ is the late time stable attractor provided   $g'(\lambda_{\phi}^{\ast}) \lambda_{\phi}^{\ast}<0$, otherwise, it is a saddle point.
\end{itemize}

 \subsection{Cosmological consequences}

As was shown in the previous subsection the autonomous systems only admits 
seven classes of critical points (some of them are actually curves) \footnote{$P_{1}^\pm$  and $P_{3}^\pm$ are ruled out. The first one because of they lead to imaginary values of dimensionless variable $x_\phi$. And the last one because of is outside of the physical phase space, representing a critical point with a negative energy density $\Omega_{\phi}<0$.}. The curves $P_{2}^\pm$ correspond to decelerated solutions, with $q=2$, where the Friedmann constraint (\ref{F1}) is dominated by the kinetic energy of the quintessence field with an equation of state of stiff type, $w_{eff}=1$. These solutions are only relevant a early times and should be unstable \cite{Copeland1998}. Unfortunately these critical points are nonhyperbolic (it has two zero eigenvalues) meaning that is not possible to obtain conclusions about its stability with the previous linear analysis. However the numerical analysis performed in the next subsection with a particular potentials confirm the previous results in 
literature.

An important result come from the stability of critical point $P_4$. This points exists if $-\sqrt{6}\leq\lambda_{\sigma}^{\ast}\leq\sqrt{6}$ and always behave as a saddle fixed point. The latter means that under certain initial conditions the orbits in the phase space will approach to this point spending some time in its vicinity before being repelled toward the attractor solution of the system. In the case of this point, as we mentioned before, if the quintessence potential fulfill the condition:
\begin{equation}\label{condiDEDM}
 \lambda_{\sigma}^{\ast}=\pm\sqrt{3}
\end{equation}
then the effective equation of state of this dark energy component would mimic pressureless fluid ($w_{eff}=0$), in other words: it will dynamically behave exactly as cold dark matter. The possibility of this dynamical characteristic impose a fine tunning over the shape of quintessence potentials and a priori there is no guarantee that all possible quintessence potentials may satisfy the above condition (\ref{condiDEDM}). Let's note that in order to obtain the lower possible dimensionality of the phase space and to studying in a relatively simple way the effects of include arbitrary quintom potentials, we have neglected  the contribution of the usual matter fields: radiation and baryonic matter in our model \footnote{See Eqs. (\ref{action}-\ref{F1}).}. As a result, a full study of important aspects, derived from realization of condition (\ref{condiDEDM}), such as: transition redshift between the decelerated and accelerated expansion phase and the clustering properties of this \textit{effective dark matter} 
are beyond the present study and will be left for a future paper.

Another important characteristics of the model is the presence of three accelerated solutions, described by critical points $P_5$, $P_6$ and $P_7$. All of them are  de Sitter solutions ($w_{eff}=-1$) dominated by the potentials of the scalar fields. As in the case of $P_4$, they behave as saddle points and, depending on the initial conditions, the orbits can evolve from the unstable fixed point ($P_{2}^\pm$ in our case) towards one or the other of the saddle points. A favorable scenario would be one in which the initial condition lead to an evolution from $P_{2}^\pm$ to the saddle point $P_4$ \footnote{we are assuming that if (\ref{condiDEDM}) is fulfilled, then quintessence field behave as the Dark Matter.} and then, the orbits tend to one of the de Sitter solutions $P_5$, $P_6$ or $P_7$ or to the late time phantom attractor ($P_8$). In terms of the cosmological evolution of the Universe, the above favorable scenario implies that the Universe started at early times from an stage dominated by the kinetic 
term of 
the quintessence, then evolve into an epoch dominated by the \textit{effective dark matter} and finally enter in the final phase of accelerated expansion. This accelerated phase can be the de Sitter solutions or a phantom dominated solution ($w_{eff}<-1$) \footnote{In fact, these models admits the possibility of having two stable solutions: a de Sitter solution ($P_7$) and a phantom solution ($P_8$), each one within their basin of attraction as was shown in previous subsection.}. This final stage of evolutions towards critical point $P_8$ is consistent  with the recent joint results from \textit{WMAP}+\textit{eCMB}+\textit{BAO}+$H_0$+\textit{SNe} \cite{Hinshaw2012} which suggest a mild preference for a dark energy equation-of-state parameter in the phantom region ($w_{eff}<-1$). 

Finally, in order to examine the stability of the nonhyperbolic  points that cannot consistently be studied via the present linear analysis, we present  a concrete example. We provide a numerical elaboration of the phase space orbits of the corresponding quintom model.

\subsection{$V(\sigma,\phi)=V_{0}\sinh^{2}(\alpha\sigma)+V_{1}\cosh^{2}(\beta\phi)$}\label{cosh}
This potential is derived, in a Friedmann-Robertson-Walker
cosmological model, from canonical quantum cosmology under determined
conditions in the evolution of our universe\footnote{This is part of a forthcoming paper.}, using the bohmian
formalism \cite{Guzman2007}. For this potential:
\begin{equation}\label{f1}
    f(\lambda_{\sigma})=-\frac{\lambda_{\sigma}^{2}}{2}+ 2 \alpha^{2}, \;\;\lambda_{\sigma}^{\ast}=\pm 2 \alpha, \;\;f'(\lambda_{\sigma}^{\ast})=-{\lambda_{\sigma}^{\ast}}
\end{equation}
and
\begin{equation}\label{f2}
    g(\lambda_{\phi})=-\frac{\lambda_{\phi}^{2}}{2}+ 2 \beta^{2}, \;\;\lambda_{\phi}^{\ast}=\pm 2 \beta, \;\;g'(\lambda_{\phi}^{\ast})=-{\lambda_{\phi}^{\ast}}.
\end{equation}
From the Table \ref{tab2} and the equation (\ref{f2}) we see that the condition
to ensure that Point $P_8$ has a four dimensional stable subspace is
always satisfied due to the opposite signs between
$\lambda_{\phi}^{\ast}$ and $g'(\lambda_{\phi}^{\ast})$. In order to
having achieved success scalar field dark matter domination era we
need that $\lambda_{\sigma}^{\ast}=\pm\sqrt{3},$ since this is the only way to have a standard transient matter dominated solution ($P_4$). Recall that for the choice $\lambda_{\sigma}^{\ast}=\pm\sqrt{3},$ the standard quintessence dominated solution mimics dark matter ($w_{eff}=0$). 
Imposing the condition $\lambda_{\sigma}^{\ast}=\pm\sqrt{3},$  we have as a degree of freedom the potential
parameter $\alpha$ that can be adjusted using (\ref{f1}). Furthermore, we impose
one of the following conditions:
\begin{equation}
    \lambda_{\sigma}^{\ast}=\sqrt{3},\;\lambda_\phi ^{\ast}\leq -\sqrt{6}, \;  1<x_\sigma<\sqrt{1+\frac{6}{\left(\lambda_\phi
   ^{\ast}\right)^2}}\nonumber
\end{equation} or
\begin{equation}
    \lambda_{\sigma}^{\ast}=\sqrt{3},\;-\sqrt{6}<\lambda_\phi ^{\ast}<0, \;  1<x_\sigma<\sqrt{2}\nonumber
\end{equation} or
\begin{equation}
    \lambda_{\sigma}^{\ast}=-\sqrt{3},\;\lambda_\phi ^{\ast}\leq -\sqrt{6}, \;-\sqrt{1+\frac{6}{\left(\lambda_\phi
   ^{\ast}\right)^2}}<x_\sigma<-1\nonumber
\end{equation} or
\begin{equation}
    \lambda_{\sigma}^{\ast}=-\sqrt{3},\;-\sqrt{6}<\lambda_\phi ^{\ast}<0, \;-\sqrt{2}<x_\sigma<-1\nonumber.
\end{equation}
to  guarantee that the stiff matter type solution of the quintom cosmology be the past-attractor.

\begin{figure}[th!]
\begin{center}
\includegraphics[scale=0.7]{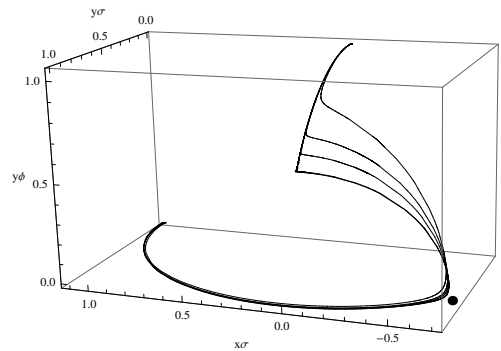}\caption{\label{fig1}Trajectories in phase space ($x_{\sigma}$, $y_{\sigma}$, $y_{\phi}$) for the potential $V(\sigma,\phi)=V_{0}\sinh^{2}(\alpha\sigma)+V_{1}\cosh^{2}(\beta\phi)$ with ($\alpha$, $\beta$): ($-\sqrt{3}/2$, $0.35$). With this parameter selection the scalar field matter dominated era ($P_4$, black point in this graphic) is a saddle, whereas, the phantom field dominated solution ($P_8$ in Table \ref{tab2}) is the late time attractor.}
\end{center}\end{figure}

\begin{figure}[th!]
\begin{center}
\includegraphics[scale=0.7]{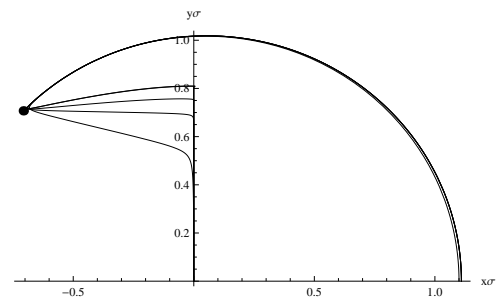}\caption{\label{fig2}Trajectories in phase space ($x_{\sigma}$, $y_{\sigma}$) with the same parameter selection of Fig. \ref{fig1}. The critical point $P_4$ represented by the black point is a scalar field matter dominated transient solution.}
\end{center}\end{figure}

\begin{figure}[h!]
\begin{center}
\includegraphics[width=6.5cm,height=5cm]{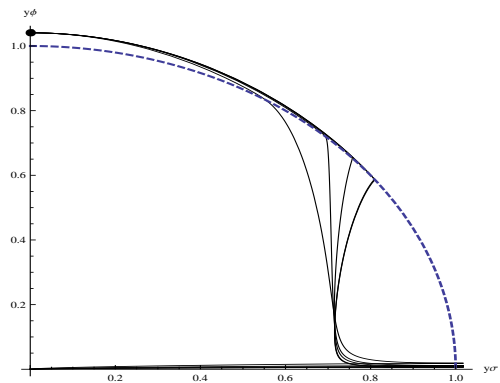}\caption{\label{fig3}Trajectories in phase space ($y_{\sigma}$, $y_{\phi}$) with the same parameter selection of Fig. \ref{fig1}. The accelerated de Sitter solution $P_7$, dashed line, is a transient era in the evolution of the Universe and the late time attractor the phantom field dominated solution, black point in this figure, allowing a crossing through the phantom divide.}
\end{center}\end{figure}

To finish this section let's discuss some numerical elaborations. In the figure \ref{fig1} are presented some trajectories in phase space ($x_{\sigma}$, $y_{\sigma}$, $y_{\phi}$) for different sets of initial conditions for potential $V(\sigma,\phi)=V_{0}\sinh^{2}(\alpha\sigma)+V_{1}\cosh^{2}(\beta\phi)$. The free parameter have been chosen to be ($\alpha$, $\beta$): ($-\sqrt{3}/2$, $0.35$). This parameter selection guarantee that point $P_4$, black point in this graphic, represents an scalar field matter (i.e., the scalar field mimicking dark matter) dominated era with a typical saddle dynamics. The late-time attractor is the phantom field dominated solution ($P_8$ in Table \ref{tab2}). In the figure \ref{fig2} are displayed some trajectories in phase space ($x_{\sigma}$, $y_{\sigma}$) with the same parameter selection as in Fig. \ref{fig1}. 
Finally, in the figure \ref{fig3} are prresented trajectories in phase space ($y_{\sigma}$, $y_{\phi}$) with the same parameter selection of Fig. \ref{fig1}. The accelerated de Sitter solution $P_7$, dashed line, is a transient era in the evolution of the Universe being the late time attractor the phantom field dominated solution, black point in this figure, allowing the crossing through the phantom divide.


\section{Conclusions}\label{ss4}
In the present paper a thorough study of the phase space of quintom
model has been undertaken. The results are valid for those
potential, without interaction between the fields, for which the
quantities
$\Gamma_{\sigma}=\frac{V_{\sigma}(\sigma)V_{\sigma}''(\sigma)}{(V_{\sigma}'(\sigma))^{2}}$
and
$\Gamma_{\phi}=\frac{V_{\phi}(\phi)V_{\phi}''(\phi)}{(V_{\sigma}'(\phi))^{2}}$
can be written as a function of the variables $\lambda_{\sigma}$ and
$\lambda_{\phi}$.

It has been found that for  $g'(\lambda_{\phi}^{\ast}) \lambda_{\phi}^{\ast}<0$, the late time attractor are always the
phantom dominated solution ($P_8$) generalizing the result
 shown in \cite{Guo2005, Cai2010} for exponential potentials (but more generally, for potential satisfying
 $\lambda_{\sigma}\approx const$ and $\lambda_{\phi}\approx const$). Otherwise, it is a saddle point. The Universe evolves from a quintessence dominated
 phase to a phantom dominated phase crossing the $w_{eff}=-1$ divide line as a transient stage \cite{Guo2006}.

 Center Manifold Theory have been  employed to analyze the stability of de Sitter solution either with phantom ($P_5$) or quintessence potential dominance ($P_6$). After deriving the evolution equation on the center manifolds and making several numerical integrations we have concluded that in both cases the corresponding de Sitter solution is unstable (saddle-like). For $P_5$ we have used an analytical argument, whereas for $P_6$ our conclusion was supported partially on analytical arguments and complemented by  numerical experimentation. 
 
 Another important issue is concerning the existence of a point, $P_4,$ corresponding to the standard quintessence dominated solution, which under certain condition on the  potential, can mimick the dark matter behavior. This feature has important cosmological consequences to address de unified description of dark matter and dark energy in a single field. The saddle type character of $P_4$ have been clearly illustrated by resorting to phase plane diagrams for the potential obtained from a canonical quantum cosmology.

\appendix
\section{Center manifold dynamics for the solution dominated by the potential energy of the phantom component $P_5$}\label{apen1}
In this section we will shown how we can apply the center manifold
theorem to study the stability of non-hyperbolic point $P_5$ corresponding to the solution dominated by the potential energy of the phantom component
\cite{perko2001differential}.
First, we restrict our attention to the domain $-\frac{3}{4}<g(0)<0$ to dealing with real eigenvalues.
The first step is to translate the
point $P_5$ ($x_{\sigma}=0$, $x_\phi=0$ $y_{\sigma}=0$,
$\lambda_{\sigma}=\mu$, $\lambda_{\phi}=0$) to the origin, where $\mu$
denotes an arbitrary value for $\lambda_{\sigma}$. The next step is
to transform the system  to its  real Jordan form:
\begin{eqnarray}
\frac{d \textbf{u}}{d N}&=&Z\textbf{u}+F(\textbf{u},\textbf{v})\label{A1}\\
\frac{d \textbf{v}}{d N}&=&P\textbf{v}+G(\textbf{u},\textbf{v})\label{A2}
\end{eqnarray}
where the square matrices $Z$, $P$ have $2$ zero eigenvalues  and $3$ eigenvalues with negative real part, respectively.
In order to do that we introduce the new variables:
\begin{eqnarray}
u_1&=&y_\sigma,\nonumber\\
u_2&=&-\sqrt{\frac{2}{3}} x_\sigma f(\mu)-\mu+\lambda _\sigma,\nonumber\\
v_1&=&\sqrt{\frac{2}{3}} x_\sigma f(\mu),\nonumber\\
v_2&=&\frac{2 \sqrt{6} g(0)
   x_\phi+\left(\sqrt{12 g(0)+9}-3\right) \lambda_\phi }{2 \sqrt{12
   g(0)+9}},\nonumber\\
v_3&=&\frac{\left(\sqrt{12 g(0)+9}+3\right) \lambda_\phi -2 \sqrt{6}
   g(0) x_\phi}{2 \sqrt{12 g(0)+9}}\end{eqnarray}
Using the above transformation, the system (\ref{A1}-\ref{A2}) is
given explicitly by:
\begin{align}
& {u_{1}}'=F_1(u_{1},u_{2},v_{1},v_{2},v_{3})\label{d1}\\
& {u_{2}}'=-\frac{3 v_1 (f(\mu+u_2+v_1)-f(\mu))}{f(\mu)}+H (u_{1},u_{2},v_{1},v_{2},v_{3})\nonumber \\ & \equiv F_2(u_{1},u_{2},v_{1},v_{2},v_{3})\label{d2}\\
& {v_{1}}'=-3v_{1}+G_1(u_{1},u_{2},v_{1},v_{2},v_{3})\label{d3}\\
& {v_{2}}'=\frac{1}{2} \left(-\sqrt{12 g(0)+9}-3\right)v_{2}+G_2(u_{1},u_{2},v_{1},v_{2},v_{3})\label{d4}\\
& {v_{3}}'=\frac{1}{2} \left(\sqrt{12 g(0)+9}-3\right)v_{3}+G_3(u_{1},u_{2},v_{1},v_{2},v_{3})\label{d5},
\end{align}
where $f'=\frac{d f}{d N},$ $F_{1}, H,G_1$ ...
$G_{3}$ are homogeneous polynomials in the coordinates $(u_{1},
u_{2},v_{1},v_{2},v_{3})$ of degree greater than $2$. Following the
standard formalism of the center manifold theory, the coordinates
which correspond to the non-zero eigenvalues ($v_1$, $v_2$, $v_3$)
can be approximated by the functions:
\begin{align}
k_{1}(u_{1},u_{2})&=a_1 u_1^2+a_2 u_1^3+a_3 u_1
u_2+a_4 u_1^2 u_2+a_5 u_2^2+\nonumber \\&+a_6
   u_1 u_2^2+a_7 u_2^3+...+ O(u_{1}^n,u_{2}^n)\label{d11}\\
k_{2}(u_{1},u_{2})&=b_1 u_1^2+b_2 u_1^3+b_3 u_1
u_2+b_4 u_1^2 u_2+b_5 u_2^2+\nonumber \\&+b_6
   u_1 u_2^2+b_7 u_2^3+...+  O(u_{1}^n,u_{2}^n)\label{d22}\\
k_{3}(u_{1},u_{2})&=c_1 u_1^2+c_2 u_1^3+c_3 u_1
u_2+c_4 u_1^2 u_2+c_5 u_2^2+\nonumber \\& +c_6
   u_1 u_2^2+c_7 u_2^3+...+  O(u_{1}^n,u_{2}^n)\label{d33}
\end{align}
 with this set of functions we can solve, to any $n$ desired degree of accurancy, the quasilinear partial differential equation for the center manifold:
\begin{equation}
Dk(\mathbf{u})\left[Z\mathbf{u}+F(\mathbf{u},k(\mathbf{u}))\right]-Pk(\mathbf{u})-G(\mathbf{u},k(\mathbf{u}))=0
\label{eq:CM}
\end{equation}
In our case: $Z=\left(
\begin{array}{cc}
 0 & 0 \\
 0 & 0
\end{array}
\right)$ and
{\small
\[ P = \left( \begin{array}{cccc}
-3 & 0 & 0 \\
 0 &\frac{1}{2} \left(-\sqrt{12 g(0)+9}-3\right) & 0 \\
 0 & 0 & \frac{1}{2} \left(\sqrt{12 g(0)+9}-3\right)
 \end{array} \right)\]
}
\[ k(\mathbf{u}) = \left( \begin{array}{c}
k_{1}(u_{1},u_2)  \\
k_{2}(u_{1},u_2)  \\
k_{3}(u_{1},u_2)
 \end{array}
\right)\]

\[ Dk(\mathbf{u}) = \left( \begin{array}{cc}
\frac{\partial k_{1}}{\partial u_{1}} & \frac{\partial k_{1}}{\partial u_{2}} \\
\frac{\partial k_{2}}{\partial u_{1}} & \frac{\partial k_{2}}{\partial u_{2}} \\
\frac{\partial k_{3}}{\partial u_{1}} & \frac{\partial k_{3}}{\partial u_{2}} \end{array}
\right) \]

{\small
\[ G(\mathbf{u},k(\mathbf{u}) )= \left( \begin{array}{c}
G_{1}\left(u_{1}, u_{2}, k_{1}(u_{1},u_2),k_{2}(u_{1},u_2),k_{3}(u_{1}, u_2)\right)  \\
G_{2}\left(u_{1}, u_{2}, k_{1}(u_{1},u_2),k_{2}(u_{1},u_2),k_{3}(u_{1}, u_2)\right)   \\
G_{3}\left(u_{1}, u_{2}, k_{1}(u_{1},u_2),k_{2}(u_{1},u_2),k_{3}(u_{1}, u_2)\right)
 \end{array} \right)\]
}

In order to solve equation (\ref{eq:CM}) we put together $Z$, $P$,
$k(u_{1},u_2)$, $Dk(u_{1},u_2)$ and $G(u_{1},u_2,k(u_{1}, u_2)),$
then we equate equal powers of $u_{1}$ and $u_2$, and in that way we
compute $k(u_{1},u_2)$. Finally we obtain, in the neighborhood of
$P_5$, the reduced system:
\begin{equation}
{\textbf{u}}'=Z\textbf{u}+F(\textbf{u},k(\textbf{u})).
\label{eq:11}
\end{equation}
The above procedure applied to equations (\ref{d1}-\ref{d5}) leads to:
\begin{equation}
v_1=\frac{1}{3}\mu f(\mu) u_1^2+\frac{1}{3} f(\mu) u_1^2 u_2+O(4); v_2=v_3=O(4).
\label{eq:fin}
\end{equation}
Neglecting the fourth order terms, the evolution equations on the center manifold are
\begin{eqnarray}
u_1'&=& -\frac{1}{2} \mu^2 u_1^3 \label{equ1}\\
u_2'&=&-u_1^2 (\mu + u_2) f(\mu)-\mu u_1^2 u_2 f'(\mu)\label{equ2}
\end{eqnarray}

For $\mu f'(\mu)+f(\mu)\neq 0,$ the orbit of \eqref{equ1}-\eqref{equ2} passing through $(u_{1 0},u_{2 0})$ is given by
\begin{equation}
\frac{u_1}{u_{1 0}}=\frac{1}{\sqrt{1+\mu^2 u_{1 0}^2N}},\; N\geq 0,
\end{equation}
\begin{equation} 
u_2=\left(\eta(\mu)+u_{20}\right) \left(\frac{u_1}{u_{10}}\right)^{\frac{2 \left(\mu f'(\mu)+f(\mu)\right)}{\mu^2}}-\eta(\mu),
\end{equation} where $\eta(\mu)=\frac{\mu f(\mu)}{\mu f'(\mu)+f(\mu)}.$ 
Then, for $ f(\mu)+ \mu f'(\mu)>0,$ the orbits approach the point with coordinates $\left(u_1=0,u_2=-\eta(\mu)\right)$ when $N\rightarrow+\infty.$ If   $ f(\mu)+ \mu f'(\mu)\leq 0,$ then, as $N\rightarrow +\infty,$ $u_1$ tends to zero and $u_2$ becomes unbounded.
Generically, the origin is not approached as $N\rightarrow +\infty,$ unless $\mu f(\mu)=0.$  

In the especial case $\mu f'(\mu)+f(\mu)=0,$  the system \eqref{equ1}-\eqref{equ2} reduces to 
\begin{eqnarray}
u_1'&=& -\frac{1}{2} \mu^2 u_1^3 \label{equ1b}\\
u_2'&=&-\mu f(\mu) u_1^2 \label{equ2b}.
\end{eqnarray}
The orbit of \eqref{equ1b}-\eqref{equ2b} passing through $(u_{1 0},u_{2 0})$ is given by
\begin{equation}
\frac{u_1}{u_{1 0}}=\frac{1}{\sqrt{1+\mu^2 u_{1 0}^2N}},\; N\geq 0,
\end{equation}
\begin{equation}
u_2=u_{20} +\ln \left[\frac{u_1}{u_{1 0}}\right]^{\frac{2 f(\mu)}{\mu}}.
\end{equation}
In this case, $u_1$ tends to zero and $u_2$ becomes unbounded.
Summarizing, for $-\frac{3}{4}<g(0)<0$, $P_5$ is unstable.

For $g(0)<-\frac{3}{4},$ there are two complex eigenvalues.
In this case, in order to obtain the real Jordan Form, we introduce the new variables
\[V_2=\frac{v_2+v_3}{2},\; V_3=\frac{v_2-v_3}{2i}.\]
Using the above transformation, the system (\ref{A1}-\ref{A2}) is
given explicitly by:
\begin{align}
& {u_{1}}'=\widetilde{F_1}(u_{1},u_{2},v_{1},V_{2},V_{3})\label{D1}\\
& {u_{2}}'=\widetilde{ F_2}(u_{1},u_{2},v_{1},V_{2},V_{3})\label{D2}\\
& {v_{1}}'=-3v_{1}+\widetilde{G_1}(u_{1},u_{2},v_{1},V_{2},V_{3})\label{D3}\\
& {V_{2}}'=-\frac{3}{2}V_2-\frac{1}{2}\sqrt{-12g(0)-9}V_3+\widetilde{G_2}(u_{1},u_{2},v_{1},V_{2},V_{3})\label{D4}\\
& {V_{3}}'=-\frac{1}{2}\sqrt{-12g(0)-9}V_2-\frac{3}{2}V_3+\widetilde{G_3}(u_{1},u_{2},v_{1},V_{2},V_{3})\label{D5}
\end{align}
where $\widetilde{F_{1}}, \widetilde{G_1}$ ...
$\widetilde{G_{3}}$ are homogeneous real polynomials in the coordinates $(u_{1},
u_{2},v_{1},V_{2},V_{3})$ of degree greater than $2$.
Usig the same procedure as before we obtain that the center manifold is given locally by the graph
\begin{equation}
v_1=\frac{1}{3}a f(a) u_1^2+\frac{1}{3} f(a) u_1^2 u_2+O(4); V_2=V_3=O(4).
\label{eq:fincomplex}
\end{equation}
Thus the dynamics on the center manifold is given by the system \eqref{equ1}-\eqref{equ2} analyzed before. Sumarizing, for $g(0)<0$, $P_5$ is unstable.

\section{Center manifold dynamics for the solution dominated by the potential energy of the quintessence component $P_6$}\label{apen2}
In this section we will shown how we can apply the center manifold
theorem to study the stability of non-hyperbolic point $P_6$ corresponding to the solution dominated by the potential energy of the quintessence component
\cite{perko2001differential}.

The first step is to translate the
point $P_6$ ($x_{\sigma}=0$, $x_\phi=0$ $y_{\sigma}=1$,
$\lambda_{\sigma}=0$, $\lambda_{\phi}=\mu$) to the origin, where $\mu$
denotes an arbitrary value for $\lambda_{\phi}$. The next step is
to transform the system  to its  real Jordan form:
\begin{eqnarray}
\dot{\textbf{u}}&=&Z\textbf{u}+F(\textbf{u},\textbf{v})\label{P6A1}\\
\dot{\textbf{v}}&=&P\textbf{v}+G(\textbf{u},\textbf{v})\label{P6A2}
\end{eqnarray}
where the square matrices $Z$, $P$ have $2$ zero eigenvalues  and $3$ eigenvalues with negative real part, respectively.
In order to do that we introduce the new variables:
\begin{eqnarray}
u_1&=&-2 \mu (y_\sigma-1) g(\mu),\nonumber\\
u_2&=&-\frac{1}{3} g(\mu) \left(\sqrt{6}
  x_\phi-2 \mu (y_\sigma-1)\right)-\mu+\lambda_\phi,\nonumber\\
v_1&=&\frac{1}{3} g(\mu) \left(\sqrt{6} x_\phi-2 \mu (y_\sigma-1)\right),\nonumber\\
v_2&=&\frac{2 \sqrt{6} f(0) x_\sigma+\left(\sqrt{9-12
   f(0)}-3\right) \lambda_\sigma }{2 \sqrt{9-12 f(0)}},\nonumber\\
v_3&=&\frac{-2 \sqrt{6} f(0) x_\sigma+\left(\sqrt{9-12
   f(0)}+3\right) \lambda_\sigma }{2
   \sqrt{9-12 f(0)}}.\end{eqnarray}
Using the above transformation, the system (\ref{P6A1}-\ref{P6A2}) is
given explicitly by:
\begin{align}
& {u_{1}}'=F_1(u_{1},u_{2},v_{1},v_{2},v_{3})\label{p61}\\
& {u_{2}}'=-\frac{3 v_1 (g(\mu+u_2+v_1)-g(\mu))}{g(\mu)}+
\frac{u_1 g(\mu+u_2+v_1)}{g(\mu)}\nonumber \\ & +H (u_{1},u_{2},v_{1},v_{2},v_{3}) \equiv F_2(u_{1},u_{2},v_{1},v_{2},v_{3})\label{p62}\\
& {v_{1}}'=-3v_{1}+G_1(u_{1},u_{2},v_{1},v_{2},v_{3})\label{p63}\\
& {v_{2}}'=\frac{1}{2} \left(-\sqrt{9-12 f(0)}-3\right)v_{2}+G_2(u_{1},u_{2},v_{1},v_{2},v_{3})\label{p64}\\
& {v_{3}}'=\frac{1}{2} \left(\sqrt{9-12 f(0)}-3\right)v_{3}+G_3(u_{1},u_{2},v_{1},v_{2},v_{3})\label{p65}
\end{align}
near the non-hyperbolic fixed point $P_6$ where $F_{1}, H,G_1$ ...
$G_{3}$ are homogeneous polynomials in the coordinates $(u_{1},
u_{2},v_{1},v_{2},v_{3})$ of degree greater than $2$. Following the
standard formalism of the center manifold theory, we obtain that the center manifold of $P_6$ is given by the graph
\begin{eqnarray}
&v_1=\frac{\mu
   u_1^2}{18 g(\mu)}+\frac{u_1^2}{12 \mu g(\mu)}+\frac{u_1^2}{9 \mu}-\frac{u_1
   u_2}{3 \mu}+O(3); \nonumber\\& v_2=O(3), v_3=O(3).
\label{eq:fin2}
\end{eqnarray}
Neglecting the third order terms, the evolution equations on the center manifold are
\begin{eqnarray}
u_1'&=& \frac{\mu u_1^2}{g(\mu)} \label{P6equ1}\\
u_2'&=&\frac{u_1 u_2}{\mu}-\frac{\left(2 \mu^2+3\right)
   u_1^2}{12 \mu g(\mu)}\label{P6equ2}
\end{eqnarray}

Let us assume $g(\mu)\neq 0, \mu\neq 0.$ Hence, the orbit of \eqref{P6equ1}-\eqref{P6equ2} passing through $(u_{1 0},u_{2 0})$ is given by
\begin{align}
&u_2=\frac{\left(\left(2 \mu^2+3\right)
   u_{10}+12 \mu^2 u_{20}-12 u_{20} g(\mu)\right)}{12 \left(\mu^2-g(\mu)\right)}\left(\frac{u_1}{u_{1 0}}\right)^{\frac{g(\mu)}{\mu^2}}+\nonumber\\&+\frac{\left(2 \mu^2+3\right) u_1}{12 \left(g(\mu)-\mu^2\right)} ,\\
&\frac{u_1}{u_{1 0}}=\frac{ g(\mu)}{g(\mu)-\mu u_{1 0} N },\; N\geq 0,
\end{align}
In order to investigate the stability of the center manifold of $P_6$ we have resorted to several numerical integrations of the system \eqref{P6equ1}-\eqref{P6equ2}. We find four typical situations that suggest that $P_6$ is unstable (saddle type).

\begin{figure}[h!]
\begin{center}
\includegraphics[scale=0.4]{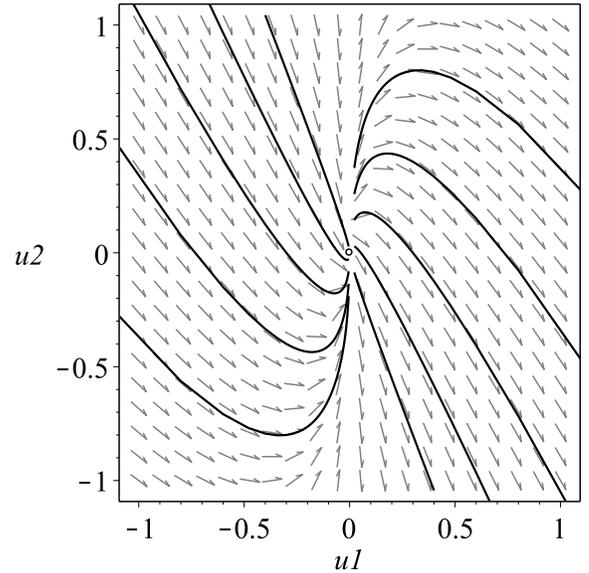}\caption{\label{fig4} Vector field in the plane ($u_1$, $u_2$) for the potential $V(\sigma,\phi)=V_{0}\sinh^{2}(\alpha\sigma)+V_{1}\cosh^{2}(\beta\phi)$. The free parameter have been chosen to be ($\alpha$, $\beta$, $\mu$): ($-\sqrt{3}/2$, $0.35$, $0.50$). In this case $g(\mu)=0.12>0$. The sign of $u_1$ is invariant. For $u_1<0$ the origin is approached as the time goes forward whereas for $u_1>0$ the orbits departs from the origin. Thus, the accelerated de Sitter solution $P_6$  is a transient era in the evolution of the Universe.}
\end{center}\end{figure}

\begin{figure}[h!]
\begin{center}
\includegraphics[scale=0.4]{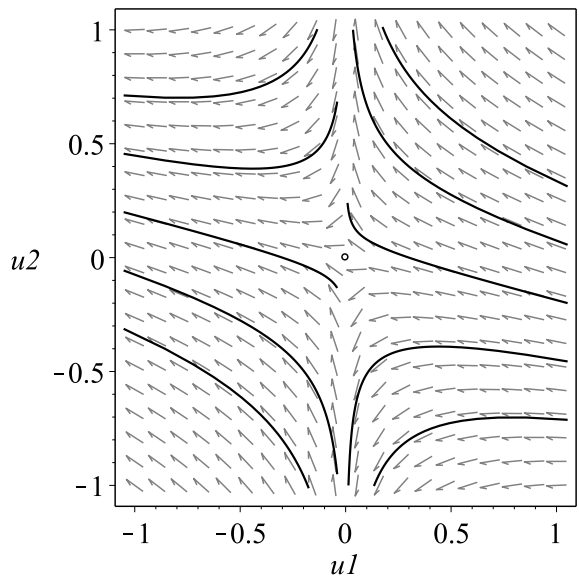}\caption{\label{fig5} Vector field in the plane ($u_1$, $u_2$) for the potential $V(\sigma,\phi)=V_{0}\sinh^{2}(\alpha\sigma)+V_{1}\cosh^{2}(\beta\phi)$. The free parameter have been chosen to be ($\alpha$, $\beta$, $\mu$): ($-\sqrt{3}/2$, $0.35$, $1.30$). In this case $g(\mu)=-0.60<0$. The origin is of saddle type. Thus the accelerated de Sitter solution $P_6$ is a transient era in the evolution of the Universe.}
\end{center}\end{figure}

In the figure \ref{fig4} is displayed the vector field in the plane ($u_1$, $u_2$) for the potential $V(\sigma,\phi)=V_{0}\sinh^{2}(\alpha\sigma)+V_{1}\cosh^{2}(\beta\phi)$. The free parameter have been chosen to be ($\alpha$, $\beta$, $\mu$): ($-\sqrt{3}/2$, $0.35$, $0.50$). In this case $g(\mu)=0.12>0$. The sign of $u_1$ is invariant. For $u_1<0$ the origin is approached as the time goes forward whereas for $u_1>0$ the orbits departs form the origin. For the choice  ($\alpha$, $\beta$, $\mu$): ($-\sqrt{3}/2$, $0.35$, $-0.50$). we have $g(\mu)=0.12>0$. The figure is similar to  \ref{fig4} with the arrows in reverse orientation. Thus, this numerical elaboration suggest that the accelerated de Sitter solution $P_6$, (the origin of coordinates), is a transient era in the evolution of the Universe for $g(\mu)>0$ irrespectively the sign of $\mu$.

In the figure \ref{fig5} is represented the vector field in the plane ($u_1$, $u_2$) for the potential $V(\sigma,\phi)=V_{0}\sinh^{2}(\alpha\sigma)+V_{1}\cosh^{2}(\beta\phi)$. The free parameter have been chosen to be ($\alpha$, $\beta$, $\mu$): ($-\sqrt{3}/2$, $0.35$, $1.30$). In this case $g(\mu)=-0.60<0$. The sign of $u_1$ is invariant. All the orbits departs from the origin. Thus the accelerated de Sitter solution $P_6$ is a transient era in the evolution of the Universe. For the choice  ($\alpha$, $\beta$, $\mu$): ($-\sqrt{3}/2$, $0.35$, $-0.30$). we have $g(\mu)=0.20>0$. The figure is similar to  \ref{fig4} with the arrows in reverse orientation. Thus, this numerical elaboration suggest that the accelerated de Sitter solution $P_6$, is a transient era in the evolution of the Universe for $g(\mu)<0$ irrespectively the sign of $\mu$.

As in the appendix \ref{apen1}, for analyzing the case of complex eigenvalues, we can introduce the new variables
\[V_2=\frac{v_2+v_3}{2},\; V_3=\frac{v_2-v_3}{2i}\] for deriving the real Jordan form of the Jacobian. The procedure is straightforward, so we won't  enter into the details here.

\section*{Acknowledgments}
This work was partially supported by PROMEP, DAIP, and by CONACyT,
M\'exico, under grants 167335 and 179881 and by  MECESUP FSM0806, from Ministerio de Educaci\'on, Chile. GL wish to thanks to his colleagues
at Instituto de F\'isica, Pontificia Universidad de Cat\'olica
de Valpara\'iso for their warm hospitality during the completion
of this work. YL is grateful to the Departamento de F\'isica and the
CA de Gravitaci\'on
y F\'isica Matem\'atica for their kind hospitality and their joint support for a
postdoctoral fellowship.

 \bibliographystyle{unsrt}
\bibliography{bib}
\end{document}